# U-Net-and-a-half: Convolutional network for biomedical image segmentation using multiple expert-driven annotations

Yichi Zhang[1], Jesper Kers[2, 3], Clarissa A. Cassol[4], Joris J. Roelofs[2, 5], Najia Idrees[6], Alik Farber[7], Samir Haroon[8], Kevin P. Daly[8], Suvranu Ganguli[8], Vipul C. Chitalia[6, 9], Vijaya B. Kolachalama[1, 10, *]

[1]Section of Computational Biomedicine, Department of Medicine, Boston University School of Medicine, Boston, MA, USA – 02118
[2]Department of Pathology, Academic Medical Center, Meibergdreef 9, 1105 AZ Amsterdam, The Netherlands
[3]Department of Pathology, Leiden University Medical Center, Albinusdreef 2, 2333 ZA Leiden, The Netherlands; Van 't Hoff Institute for Molecular Sciences (HIMS), University of Amsterdam, Science Park 904, 1098 XH Amsterdam, The Netherlands
[4]Arkana laboratories, Little Rock, AR, USA – 72211
[5]Amsterdam Cardiovascular Sciences, University of Amsterdam, The Netherlands
[6]Renal Section, Department of Medicine, Boston University School of Medicine, Boston, MA, USA – 02118
[7]Department of Surgery, Boston University School of Medicine, Boston, MA, USA – 02118
[8]Department of Radiology, Boston University School of Medicine, Boston, MA, USA – 02118
[9]Veterans Affairs Boston Healthcare System, Boston MA 02118, USA
[10]Department of Computer Science, College of Arts & Sciences, Boston University, MA, USA – 02215; Faculty of Computing & Data Sciences, Boston University, Boston, MA, USA – 02215

**\*Corresponding author:**
Vijaya B. Kolachalama, PhD; 72 E. Concord Street, Evans 636, Boston, MA - 02118
Email: vkola@bu.edu; Phone: (+1) 617-358-7253; Twitter: @vkola_lab
ORCID: https://orcid.org/0000-0002-5312-8644

**Short title:** U-Net-and-a-half

**Abstract**

Development of deep learning systems for biomedical segmentation often requires access to expert-driven, manually annotated datasets. If more than a single expert is involved in the annotation of the same images, then the inter-expert agreement is not necessarily perfect, and no single expert's annotation can precisely capture the so-called ground truth of the regions of interest on all images. Also, it is not trivial to generate a reference estimate using annotations from multiple experts. Here we present a deep neural network, defined as *U-Net-and-a-half*, which can simultaneously learn from annotations performed by multiple experts on the same set of images. *U-Net-and-a-half* contains a convolutional encoder to generate features from the input images, multiple decoders that allow simultaneous learning from image masks obtained from annotations that were independently generated by multiple experts, and a shared low-dimensional feature space. To demonstrate the applicability of our framework, we used two distinct datasets from digital pathology and radiology, respectively. Specifically, we trained two separate models using pathologist-driven annotations of glomeruli on whole slide images of human kidney biopsies (n=10 patients), and radiologist-driven annotations of lumen cross-sections of human arteriovenous fistulae obtained from intravascular ultrasound images (n=10 patients), respectively. The models based on *U-Net-and-a-half* exceeded the performance of the traditional U-Net models trained on single expert annotations alone, thus expanding the scope of multitask learning in the context of biomedical image segmentation.

**Introduction**

The application of convolutional neural networks, especially the well-known U-Net architecture [1], has given rise to several promising applications focused on biomedical segmentation in the fields of radiology and digital pathology [2]. The U-Net consists of a contracting path (or an encoder) to capture the context within an image (or an image patch), and a symmetric expanding path (or a decoder) that enables precise localization. This architecture has shown to outperform more traditional techniques such as the ones using sliding-window convolutional networks for image segmentation [3]. Subsequently, architectures that were built on top of the U-Net framework such as U-Net++ [4], H-denseunet [5], MultiResUNet [6], and many others demonstrated a range of improved performances on various biomedical image segmentation tasks [7]. Most of these extensions leverage novel implementations such as using short skip connections as well as nested and dense skip connections with or without deep supervision to effectively capture finer aspects of the image features, resulting in improved computational efficiency and model performance. Extension of the U-Net architecture for volumetric images was also proposed (3D U-Net) [8], where all the 2D operations were replaced with 3D counterparts. All these U-Net-based deep learning frameworks rely on expert-driven annotations of the regions of interest (ROI), which are typically used as reference estimates for model training and testing. As such, the annotation task is a manual, labor intensive process and prone to human error, thus leading to imperfect precision, especially when the expert annotates the ROIs across several images. If multiple experts are involved in annotations on the same set of images, then the errors in precision are likely to increase further. Moreover, choosing a reference estimate from multiple expert-driven annotations for training and testing the neural network is also not trivial. The existing U-Net and U-Net-based architectures are not equipped to learn the image segmentation task in such scenarios containing independent annotations on the same images from multiple experts.

We developed a modified U-Net architecture termed as '*U-Net-and-a-half*', which can simultaneously learn from multiple annotations generated by different experts on the same set of images. The basic architecture involves a single encoder and multiple decoders that allow simultaneous learning from image masks obtained from multiple expert-driven annotations, along with a shared low-dimensional feature space. Previous work using multiple encoder-decoder configurations was done on scene understanding, machine translation and image captioning tasks. For example, Kuga and colleagues performed multi-task learning by combining multimodal data with a shared latent representation between multiple encoder-decoder networks [9]. Baier et al., leveraged fused hidden representations of multiple encoder networks using an attention mechanism for modeling distributed sensor networks [10]. Recently, a two-stage convolutional encoder-decoder network was proposed for topology optimization [11]. Also, Eigen et al., combined the outputs of several expert networks using a multiple decoder architecture, where a gating network was trained to map each input to a distribution over the experts [12]. In the context of biomedical images, researchers have used fused encoder-decoder configurations to improve segmentation tasks, where each encoder-decoder network was used to tackle a specific task such as region- or edge-based segmentation [13]. On the other hand, multiple encoders with a single decoder were used, where multimodal brain images were fused together to improve biomedical segmentation [14-16]. Several other methodologies for biomedical segmentation tasks can be found in the literature [17-20]. We note these examples as motivation to the proposed work, where our objective was to leverage independent annotations from the experts, and simultaneously learn from them by not prioritizing one expert's annotations over the other.

The goal of this study is to develop a computational pipeline that can learn the annotations of experts with different experiences and generate a generalizable framework for biomedical segmentation. To achieve this goal, we sought the expertise of practicing radiologists and pathologists to independently annotate the regions of interest in digital images. We selected glomerular segmentation on digitized human kidney biopsies and luminal segmentation from intravascular ultrasound (IVUS) images of human arteriovenous fistulae obtained from patients with end-stage kidney disease (ESKD), to demonstrate the applicability of *U-Net-and-a-half* for image segmentation in the fields of digital pathology and radiology, respectively. Our underlying hypothesis is that *U-Net-and-a-half* can simultaneously learn from the similarities as well as disagreements between multiple expert-driven annotations and capture needed information for the segmentation task. Such a learning task would essentially generate predictions that serve as the best compromise between the different expert annotations, leading to a more generalizable framework than the traditional U-Net model trained on single expert-driven annotations alone. We evaluated the performance of *U-Net-and-a-half* using various metrics and compared them with the models based on traditional U-Net architecture on both the glomerular and IVUS imaging datasets.

**Materials and methods**

*Pathology imaging dataset*

We obtained de-identified whole slide images (WSIs) of Periodic acid-Schiff (PAS)-stained kidney biopsies of patients (n=10) who underwent a kidney transplant biopsy at the Academic Medical Center (AMC), a hospital affiliated with the University of Amsterdam (**Table 1 & Figure 1**). Renal biopsy as well as patient data collection, staining and digitization followed protocols approved by the Institutional Review Board at AMC (Study number: W19_215 #19.260; Informed consent was waived). The anonymized WSIs were generated by digitizing using the Philips IntelliSite Ultrafast WSI scanner at 20x apparent magnification (40x digital magnification) with 0.25 µm per pixel resolution and saved in bigTIFF format. All the WSIs were uploaded to a secure, web-based software (PixelView, deepPath, Inc.). A manual quality check was performed on all the WSIs. This step ensured there were no artifacts on the selected WSI regions such as air bubbles, folding, compressing, tearing, over- or under-staining, stain batch variations, knife chatter and thickness variances. Since all the WSIs had multiple cores, we were able to select a core that had no quality issues in all cases. The selected portion on each WSI was then used for glomerular annotation and subsequent analysis. Two nephropathologists (J.J.R and C.A.C) accessed the WSIs using the software and independently annotated all the glomeruli. Each WSI was opened on an iPad and an Apple Pencil was used to carefully capture the Bowman's capsule. The annotated WSIs were downloaded on a local workstation and image patches of size 512×512 pixels, with 50% overlap between adjacent patches were created. The cropped image patches containing renal tissue within at least 35% or more pixels were considered for model development, resulting in a total of 12,760 image patches that were used for model development.

*Radiological imaging dataset*

We obtained de-identified intravascular ultrasound (IVUS) images of arteriovenous fistulae in patients (n=10), admitted to Boston Medical Center (BMC) (**Table 2 & Figure 2**). Imaging procedure as well as patient data collection followed standard protocols approved by the Institutional Review Board at Boston University Medical Campus (Study number: H-37396 & H-26367). All these patients had end-stage kidney disease (ESKD) and were referred to the BMC interventional radiology suite for evaluation of their dialysis access. The IVUS scan containing several cross-sectional images per patient was uploaded to PixelView. A manual check was performed by a radiologist (K.P.D), and images with no artifacts were selected for next steps.

The lumen cross-sections on the selected images were independently annotated by the two radiologists (S.H and K.P.D). Care was taken to outline the inner wall of the arteriovenous vasculature. A total of 249 IVUS images were annotated and used for model development.

*Deep learning architecture*

Our proposed deep learning framework, referred to as *U-Net-and-a-half*, is designed to simultaneously learn from annotations performed by more than a single expert on the same set of images (**Figure 3**). The basic architecture consists of an encoder that takes the image as the input and generates a feature space, which is shared with all the decoders, where each decoder attempts to capture the annotation expertise of an expert. While in principle, this idea can apply to multiple expert-driven annotations, we demonstrated its feasibility using WSIs and IVUS images annotated by two independent experts, respectively.

The encoder was built by combining an existing backbone network (ResNet50), which consists of convolutional blocks containing a convolutional layer with batch normalization and a rectified linear unit activation function. We used maxpooling for downsampling and set the initial number of filters to 64. While we utilized ResNet50 for this task, other networks such as VGG, DenseNet or the original U-Net can also be used. The encoder connects with each decoder via the shared feature space, and each decoder comprises skip connections, which are like the original U-Net architecture. Each decoder learns from the corresponding image masks and the final prediction was generated by aggregating the outputs from both the decoders. In essence, this framework's design is motivated by a reinforcement learning task. The shared features of the backbone network were learned by a sequential loss back propagation from two decoders. If $O_1$ and $O_2$ are the separated outputs from two decoders and the reference annotations are $A_1$ and $A_2$, then the loss initialized at decoder $D_1$ was calculated from ($O_1 + O_2$) against $A_1$ and the loss initialized at decoder $D_2$ was calculated from ($O_1 + O_2$) against $A_2$. Specifically, cross entropy (CE) loss functions with respect to each decoder were combined to generate a weighted total loss defined as

$$\mathcal{L}_{hybrid}(p, y_1, y_2) = -w \sum_{i}^{n} p^i * \log(y_1^i) - (1-w) \sum_{i}^{n} p^i * log(y_2^i) \quad (1)$$

where *p* is the prediction, and $y_1$ and $y_2$ are corresponding annotations, respectively. To boost performance, the losses during back propagation in each training epoch were dynamically weighted, where the weight *w* was initialized based on the Dice coefficient and the intersection of union of the current pair of reference annotations. Note that while it is reasonable to assign

*w*=0.5, which assumes equal importance on both the expert annotations, it could be advantageous for the model to learn segmentation in a more generalizable fashion if the parameter can be learned dynamically during model training. If the annotation expertise is similar, then *w* might end up being 0.5 but if one expert's annotations are more variable, then the model might learn these differences via assigning a different weight.

In addition to CE loss, we also incorporated a hybrid focal loss, a modification of the CE loss to replace the weighted CE loss used in WSI model training only. Typically, loss functions in the training are not restricted but due to the large number of patches in the WSIs dataset and the nature of object detection, the introduction of the focal loss increased the training efficiency as well as to prevent overfitting, where the number of epochs reduced from 50 to 25 for convergence in our initial set of experiments. The loss is defined as

$$\mathcal{L}_{focal} = w * cw * (1 - e^{-CE_1})^\gamma * CE_1 + (1 - w) * cw * (1 - e^{-CE_2})^\gamma * CE_2 \quad (2),$$

where *cw* is the weight of a positive class in an imbalanced dataset, in this case, 0.25 was used as the portion of patches containing glomeruli. Gamma is a parameter that adjusts the speed of convergence.

All input images were resized to a size of 224×224 pixels. We used on-the-fly data augmentation comprising: (1) random rotation of the images ranging from -180 degrees to +180 degrees on y axis, (2) random mirror flipping with a probability of 0.5 on x and y axes, (3) random cropping with subsequent resizing with a probability of 0.5, and (4) random adjusting brightness in a range of 0.4 to 1.6 for IVUS scans and random adjusting of hue and contrast for WSI patches to account for variability observed during data collection.

We developed our framework using PyTorch and model training was performed on a GPU workstation containing a NVIDIA's GeForce RTX 3090 graphic card with a 24-Gb GDDR6X memory. Model training took less than 4 hours on WSIs or 10 minutes on IVUS scans to reach convergence, whereas prediction on the test images was an efficient process taking 60 seconds per WSI containing approximately 2,500 patches, and 5 seconds on the entire IVUS scan containing approximately 200 images.

*Performance metrics*

We used metrics including the Dice coefficient (D), intersection of union (IoU), and core Dice coefficient (cD) to evaluate the agreement between the pathologists on the glomerular annotations and between the radiologists on the IVUS annotations, respectively. For example, the Dice coefficient between pathologist 1 and pathologist 2 is denoted as $D^{path}$ and the Dice coefficient between radiologist 1 and radiologist 2 is denoted as $D^{rad}$. The performance of *U-Net-and-a-half* was also computed using D, IoU and cD metrics. For comparison, we also computed D, IoU and cD on the traditional U-Net models trained on a single expert's annotations and used these models to test on the other expert's annotations. For example, the Dice coefficient evaluated on the U-Net model trained on the annotations of expert "*i*" and tested on the annotations by expert "*j*" is denoted as $D_i^j$. The performance metrics evaluated on *U-Net-and-a-half* are denoted with the subscript "*unaah*" and with a superscript "*i*" indicating the annotations of the expert "*i*" on which *U-Net-and-a-half* was tested upon.

*Data sharing*

Computer scripts and processed data are made available on GitHub (https://github.com/vkola-lab/unaah). Raw data from AMC and BMC can be obtained upon request and subjected to institutional approval.

**Results**

The inter-annotation agreement between the pathologists on the WSIs was excellent ($D^{path}$=0.9878 and $cD^{path}$=0.9352), indicating consistency between the expert annotations. Note that $D^{path}$ was measured by considering all the image patches that have the tissue, and $cD^{path}$ was computed on only those tissue image patches that have annotations by at least one pathologist. It is not surprising to observe that $cD^{path}<D^{path}$, given that 81% of patches in this imbalanced dataset had no annotations, and this would have contributed to a higher value of $D^{path}$. Similarly, the $IoU^{path}$=0.9802 computed on all the tissue image patches was higher than $IoU^{path\_nobk}$=0.8953, where $IoU^{path\_nobk}$ denotes the metric computed only on those image patches that contained annotations by at least one pathologist. With respect to the IVUS images, the agreement between the radiologists was also very high ($D^{rad}$=0.9646 and $IoU^{rad}$=0.7851). It must be noted $D^{rad}$ and $cD^{rad}$ refers to the same metric in the context of IVUS images because all of them contained images with at least one expert annotation. Keeping the annotation expertise of the clinicians aside, we contend that the lower image resolution on IVUS images (relative to WSIs) could have contributed to generating agreement with lower values of Dice coefficient ($D^{path}>D^{rad}$) and intersection over union ($IoU^{path}>IoU^{rad}$), between the radiologist annotations.

*U-Net-and-a-half* accurately learned the annotations on the pathology (**Table 3 & Figure 4**), and radiology (**Table 4 & Figure 5**) images, at least as indicated by various metrics on each dataset. Moreover, it was interesting to see that *U-Net-and-a-half* performed better than the traditional U-Net models that were trained on a single expert but tested on the other expert's annotations. This result indicates a strong degree of model generalizability. On the other hand, it was not surprising to see the traditional U-Net trained on a single expert's annotations performing better when tested on the images annotated by the same expert but poorly when tested on the images annotated by the other expert. Similar inferences can be made when other metrics such as IoU and cD are used to compare *U-Net-and-a-half* with the traditional U-Net models.

Visually, predictions made by *U-Net-and-a-half* show apparent differences between what the single encoder and two-decoder architecture learned from two different reference estimates (i.e., annotations by two experts), and the U-Net models based on annotated images from a single expert alone (**Figures 4 & 5**). Clearly, the differences between the expert annotations on the glomerular images look subtle when compared with the apparent differences observed between the expert annotations on the IVUS images. It is also interesting to note how the predictions of the ROIs for the IVUS images appear to be a compromise between the two expert annotations compared to the WSIs. Additional WSI and IVUS cases that showcase the performance of *U-Net-and-a-half* can be found in the supplement (**Figures S1 & S2**).

**Discussion**

In this work, we developed a novel deep learning architecture (*U-Net-and-a-half*) for biomedical image segmentation, which can simultaneously learn from annotations generated by multiple experts. The architecture which was based on the traditional U-Net framework contained a single encoder, multiple decoders, and a shared feature space. We demonstrated the applicability of *U-Net-and-a-half* on two unique test cases involving pathology slides and radiology images (WSIs and IVUS images, respectively), showcasing the utility of developing generalizable frameworks for biomedical segmentation. The strength of *U-Net-and-a-half* is in its ability to capture the annotation patterns from multiple experts by simultaneously minimizing the losses from both the decoders and backpropagating the net loss to the shared feature space and then to the encoder. This strategy allowed us to achieve more accurate image segmentation than the standalone U-Net architecture built on single expert annotations and tested on the other expert data, thus creating a compromise between the two experts. Thus, our findings demonstrate innovation at the heart of digital pathology and radiology, simultaneously contributing novel insights to the field of computer vision while also expanding the scope of deep neural networks for biomedical segmentation.

The scientific community recognizes the important role of experts in annotating regions of interest (ROIs) and the tedious amount of work needed to create data for training and testing image segmentation models. To minimize their burden, several novel techniques were proposed for segmentation that are based on weakly supervised learning [21-35]. The underlying concept is to let the expert annotate ROIs on a few cases and develop advanced deep learning techniques that can learn from those small set of cases. This strategy has been quite successful recently as researchers have demonstrated that weakly supervised methods can approach the performance of fully supervised systems [36]. However, there is limited work related to image segmentation when more than a single expert is involved in image annotation. We specifically tackled this aspect by creating a framework that can simultaneously learn from multiple expert annotations. While we demonstrate the advantage of using *U-Net-and-a-half* for biomedical segmentation using two experts on each test case (i.e., single encoder and two decoders), this framework can be easily extended to learn from multiple experts. The basic idea would be to have a unique decoder for each expert's annotations, and backpropagate the overall loss in a similar fashion to the shared feature space during model development.

We selected the glomerular annotation on WSIs and lumen annotation on IVUS images because both have approximate circular shapes. While the annotation task as such was to capture the circular shaped boundaries on both tasks, the expert's attention on the WSIs was focused mainly by reviewing what's on the inside of the Bowman's capsule. Interestingly, on the other hand, the experts annotating the IVUS images were mainly focused to capture the lumen boundary by focusing on the outside of the lumen. These aspects simply underscore the subtle differences in the expert's approach to the annotations, while also highlighting the shape similarities between the captured ROIs on both the imaging datasets. Importantly, these important anatomic structures allowed us to demonstrate the proof of principle that *U-Net-and-a-half* can learn from multiple expert annotations.

We acknowledge the following limitations in our study. *U-Net-and-a-half* learns from annotations derived from multiple experts and attempts to create a compromise between them. This strategy will likely work better than single U-Net architectures when no expert annotation resembles the so-called ground truth, which is probably the case in most real-world scenarios. We must also note that implementation of *U-Net-and-a-half* requires manual annotations from more than one expert on the same set of images, which implies that we are increasing the experts' time and effort on annotations. We recommend using this strategy in several real-world scenarios, especially when the inter-expert annotations are less likely to overlap with each other. As such, our modeling framework is contributing to the growing literature on creating generalizable deep learning algorithms at the cost of expert availability. Future work can focus on combining weakly supervised learning with our proposed multitask learning strategy to minimize the experts' time on annotations, while simultaneously creating more generalizable image segmentation frameworks.

In conclusion, we demonstrated the effectiveness of simultaneously capturing the annotations of multiple experts using a single deep neural network, and this strategy turned out to be superior to using independent U-Net models for each expert. Our architecture also serves as an efficient template to bring consensus among annotations from multiple experts. Importantly, such methods hold the potential to give reproducible findings, especially when experts are involved in the annotation of non-trivial regions of interest on images and when there is limited consensus among them. Future studies can focus on evaluating the additional benefit of multitask learning for biomedical segmentation.

**Funding**

This work is supported by grants from the Karen Toffler Charitable Trust, the NIDDK Diabetic Complications Consortium grant (Subaward:32307-93 from U24-DK115255), a Strategically Focused Research Network (SFRN) Center (20SFRN35460031) from the American Heart Association, the Hariri Institute for Computing and Computational Science & Engineering at Boston University, the Dutch Kidney Foundation (17OKG23), and NIH (R21-CA253498, R21-DK119740, R01-HL132325 and R25-DK128858).

**Conflicts of interest**

None

## Tables and figures

**Table 1: Pathology imaging study population.** Whole slide images of kidney biopsies were obtained on 10 patients admitted to Academic Medical Center, a hospital affiliated with the University of Amsterdam. Several data including their demographic and clinical characteristics are reported in the table. IQR = interquartile range; IFTA = interstitial fibrosis and tubular atrophy; HLA = human leukocyte antigen; DSA = donor-specific antibodies.

| Characteristic | Value | Units |
|---|---|---|
| Number of patients | 10 | - |
| Age, median (IQR) | 43 (29 - 59) | years |
| Male | 40 | % |
| BMI, median (IQR) | 23 (19 – 25) | kg/m$^2$ |
| Prior dialysis | 70 | % |
| Diabetes mellitus | 10 | % |
| - Biopsy days after transplantation, median (IQR) | 333 (116 – 1127) | days |
| - Rejection in biopsy | 50 | % |
| - Banff minimum biopsy (7 glomeruli, 1 artery) | 100 | % |
| - Global glomerulosclerosis, median (IQR) | 5 (0 – 9) | % of glomeruli |
| - Banff g-score, median (IQR) | 1 (0 – 3) | - |
| - Banff cg-score, median (IQR) | 0 (0 – 2) | - |
| - Banff mm-score, median (IQR) | 0 (0 – 0) | - |
| - Banff i-score, median (IQR) | 0 (0 – 0) | - |
| - Banff t-score, median (IQR) | 0 (0 – 0) | - |
| - Banff total inflammation, median (IQR) | 15 (10 – 30) | % |
| - Banff IFTA, median (IQR) | 15 (10 – 30) | % |
| - Banff i-IFTA-score, median (IQR) | 3 (1 – 3) | - |
| - Banff ptc-score, median (IQR) | 0 (0 – 3) | - |
| - Banff v-score, median (IQR) | 0 (0 – 0) | - |
| - Banff cv-score, median (IQR) | 1 (0 – 1) | - |
| - Banff ah-score, median (IQR) | 1 (0 – 2) | - |
| - C4d positive | 30 | % |
| - Known anti-HLA DSA (Luminex) | 30 | % |
| Death-censored graft failure | 50 | % |
| Deceased | 30 | % |

**Figure 1: Representative WSI with glomerular annotations.** A sample WSI of a human kidney needle biopsy processed with Periodic acid–Schiff stain is shown. The inset images include a selected WSI region without the glomerular annotations as well as the same WSI region independently annotated by two different pathologists. The glomerular annotations are shown in green and blue, where green color indicates a normal glomerulus, and the blue indicates a partial or a fully sclerosed glomerulus.

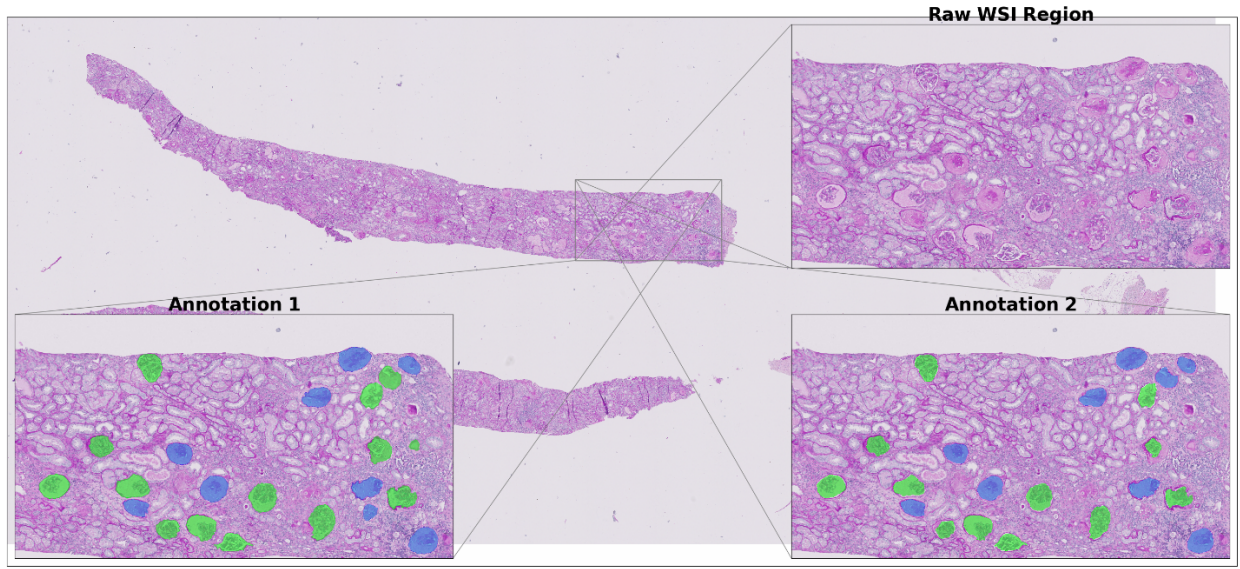

**Table 2: Radiological imaging study population.** Intravascular ultrasound images were obtained on the arteriovenous fistulae on 10 patients with end-stage kidney disease, who were referred to Boston Medical Center. Several data including their demographic and clinical characteristics are reported in the table. ESKD: End-stage kidney disease; CHF: Congestive heart failure; CAD: Coronary artery disease; PAD: Peripheral artery disease; AVF: Arteriovenous fistula; AVG: Arteriovenous graft.

| **Demographics** | **Variable** | **n** | **%** |
|---|---|---|---|
| Sex | Female | 6 | 60 |
|  |  |  |  |
| Median age |  | 69.5 years | |
|  |  |  |  |
| Race |  |  |  |
|  | Black | 8 | 80 |
|  | Hispanic | 2 | 20 |
| Comorbidities |  |  |  |
|  | Smoking | 5 | 50 |
|  | CHF | 7 | 70 |
|  | CAD | 6 | 60 |
|  | PAD | 2 | 20 |
|  | HLD | 5 | 50 |
|  | Thromboembolism | 4 | 40 |
|  |  |  |  |
| Cause of ESKD |  |  |  |
|  | Diabetes | 6 | 60 |
|  | Hypertension | 9 | 90 |
|  | Lupus nephritis | 1 | 10 |
|  | Cancer | 1 | 10 |
|  | Other/Unknown | 1 | 10 |
|  |  |  |  |
| AVF vs AVG | AVF | 6 | 60 |
|  | AVG | 4 | 40 |

**Figure 2: Representative IVUS images with lumen annotations.** The first column shows three different cross-sectional images of the arterial vessel at different locations. The second and third columns show the annotations (blue and green) performed by the radiologists on the same cross-sectional images, respectively.

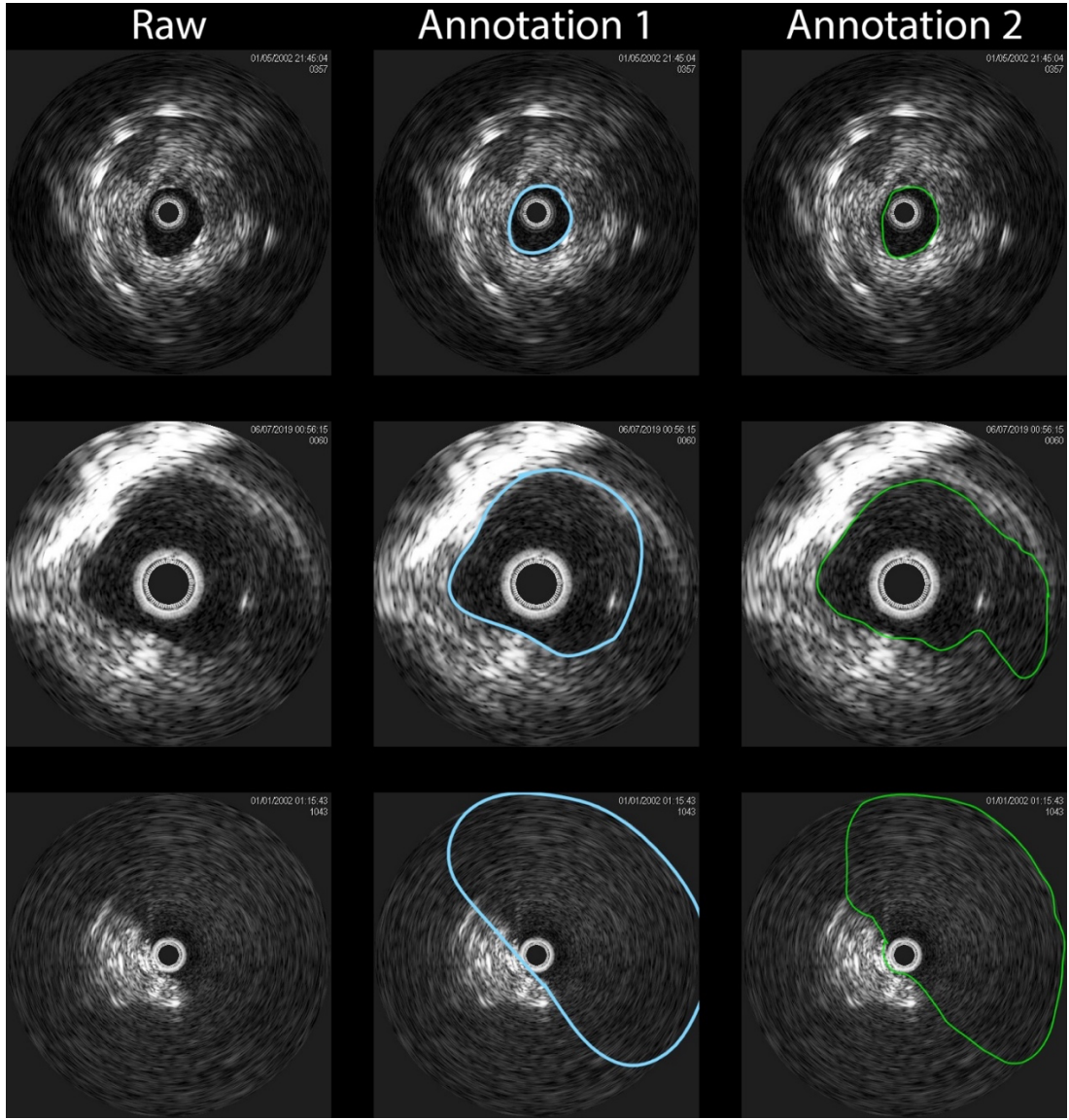

**Figure 3: Deep learning architecture.** The schematic shows the configuration of *U-Net-and-a-half*, a deep learning framework that attempts to learn from multiple annotations on the same image by different experts. Primarily, *U-Net-and-a-half* takes a single input image and predicts the region of interest by learning from two different annotations on the same image. *U-Net-and-a-half* contains a shared encoder connected with two decoders, each trying to learn from a single expert's annotation.

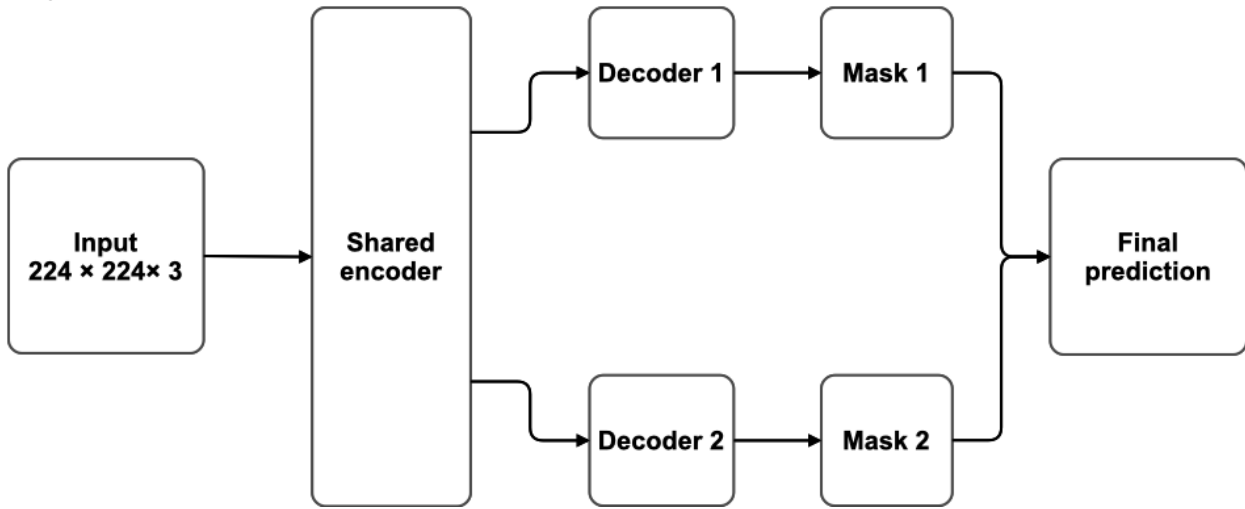

**Table 3: Model performance on the WSI dataset.** Three different metrics (Dice coefficient (D), Core Dice coefficient (CD) and intersection over union (IoU)) are reported for each model result, all on the testing data that was not used during model training. Note that the IoU metric was computed on image patches that had annotations by at least one pathologist or when the model predictions were indicating the presence of glomeruli. For Annotation 1, UNet 1 indicates a model trained and tested on the annotations by pathologist 1, UNet 2 indicates a model trained on annotations by pathologist 2 but tested on annotations by pathologist 1, and UNaah reports results trained on both the pathologists' annotations and tested on the annotations by pathologist 1. For Annotation 2, UNet 1 indicates a model trained on the annotations by pathologist 1 but tested on annotations by pathologist 2, UNet 2 indicates a model trained and tested on annotations by pathologist 2, and UNaah reports results trained on both the pathologists' annotations and tested on the annotations by pathologist 2.

| Annotation | Model | Dice Coefficient (D) | Core Dice Coefficient (cD) | Intersection over Union (IoU) |
|---|---|---|---|---|
| 1 | UNet 1 | 0.9867±0.005 ($D_1^1$) | 0.9410±0.02 ($cD_1^1$) | 0.7768±0.01 ($IoU_1^1$) |
| | UNet 2 | 0.9858±0.003 ($D_2^1$) | 0.9254±0.02 ($cD_2^1$) | 0.7183±0.03 ($IoU_2^1$) |
| | UNaah | 0.9874±0.005 ($D_{unaah}^1$) | 0.9397±0.02 ($cD_{unaah}^1$) | 0.7573±0.02 ($IoU_{unaah}^1$) |
| 2 | UNet 1 | 0.9834±0.006 ($D_1^2$) | 0.9179±0.02 ($cD_1^2$) | 0.7604±0.01 ($IoU_1^2$) |
| | UNet 2 | 0.9886±0.004 ($D_2^2$) | 0.9435±0.01 ($cD_2^2$) | 0.7846±0.02 ($IoU_2^2$) |
| | UNaah | 0.9869±0.003 ($D_{unaah}^2$) | 0.9316±0.01 ($cD_{unaah}^2$) | 0.7758±0.02 ($IoU_{unaah}^2$) |

**Figure 4: Model performance on a sample WSI patch.** In the first column, a representative WSI patch is shown with a glomerulus. In the second column, annotations performed by the two pathologists highlighting the glomerulus are shown. The third column shows the performance of the UNet 1 on annotation 1, UNaah that learned from both the annotations and UNet 2 on annotation 2.

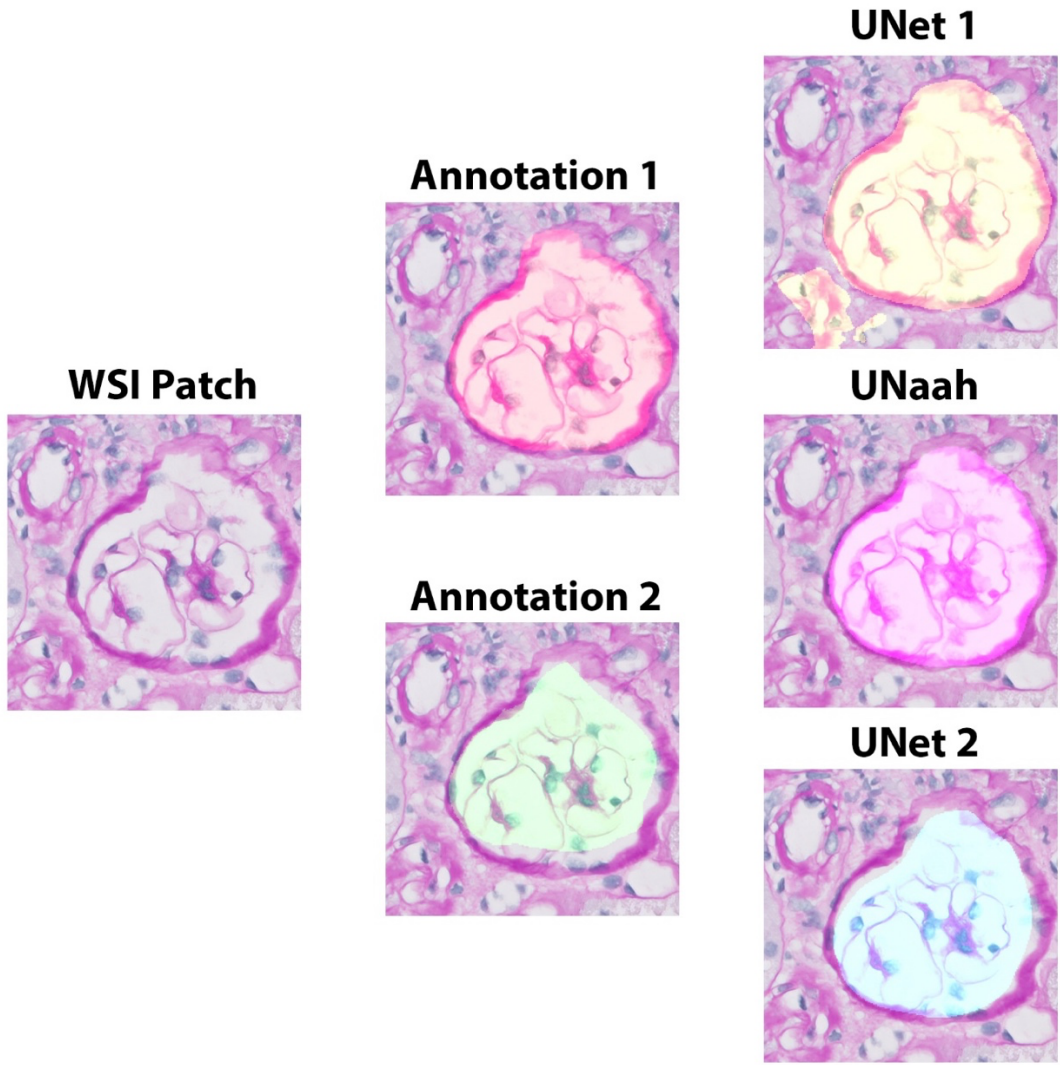

**Table 4: Model performance on the IVUS dataset.** Two different metrics (Dice coefficient (D) and intersection over union (IoU)) are reported for each model result, all on the testing data that was not used during model training. For Annotation 1, UNet 1 indicates a model trained and tested on the annotations by radiologist 1, UNet 2 indicates a model trained on annotations by radiologist 2 but tested on annotations by radiologist 1, and UNaah reports results trained on both the radiologists' annotations and tested on the annotations by radiologist 1. For Annotation 2, UNet 1 indicates a model trained on the annotations by radiologist 1 but tested on annotations by radiologist 2, UNet 2 indicates a model trained and tested on annotations by radiologist 2, and UNaah reports results trained on both the radiologists' annotations and tested on the annotations by radiologist 2.

| Annotation | Model | Dice Coefficient (D) | Intersection over Union (IoU) |
|---|---|---|---|
| 1 | UNet 1 | 0.9530±0.02 ($D_1^1$) | 0.8482±0.05 ($IoU_1^1$) |
| 1 | UNet 2 | 0.9472±0.03 ($D_2^1$) | 0.8322±0.04 ($IoU_2^1$) |
| 1 | UNaah | 0.9557±0.02 ($D_{unaah}^1$) | 0.8561±0.03 ($IoU_{unaah}^1$) |
| 2 | UNet 1 | 0.9596±0.02 ($D_1^2$) | 0.8600±0.03 ($IoU_1^2$) |
| 2 | UNet 2 | 0.9669±0.02 ($D_2^2$) | 0.8881±0.03 ($IoU_2^2$) |
| 2 | UNaah | 0.9691±0.01 ($D_{unaah}^2$) | 0.8943±0.02 ($IoU_{unaah}^2$) |

**Figure 5: Model performance on a sample IVUS image.** In the first column, a representative IVUS image is shown. In the second column, annotations performed by the two radiologists highlighting the arterial cross section are shown. The third column shows the performance of the UNet 1 on annotation 1, UNaah that learned from both the annotations and UNet 2 on annotation 2.

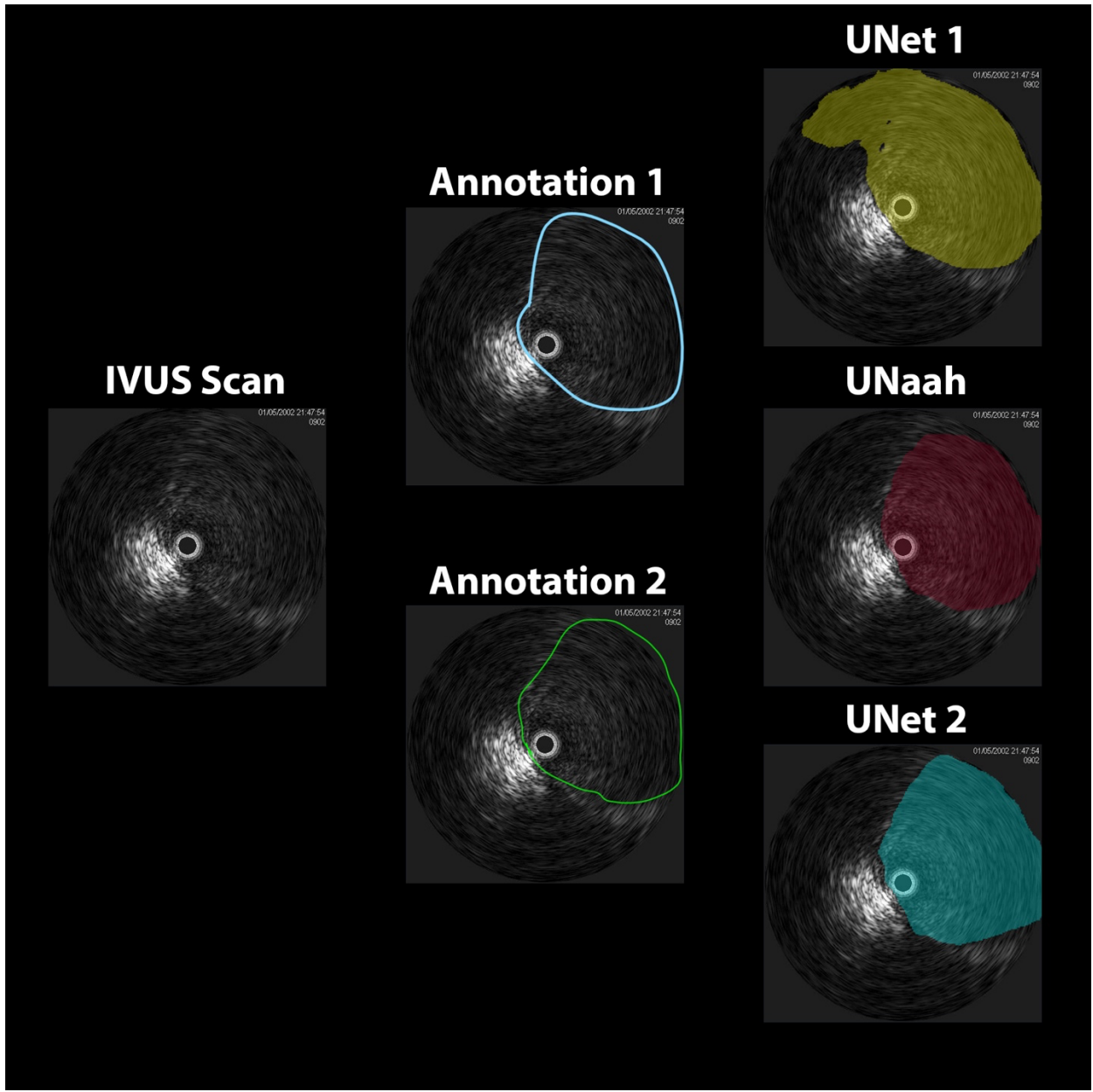

**Supplementary material**

**Figure S1: Model performance on sample WSIs.** In each sub-figure, the first column shows a representative WSI patch, and the second column shows annotations performed by the two pathologists highlighting the glomerulus. The third column shows the performance of the UNet 1 on annotation 1, UNaah that learned from both the annotations and UNet 2 on annotation 2.

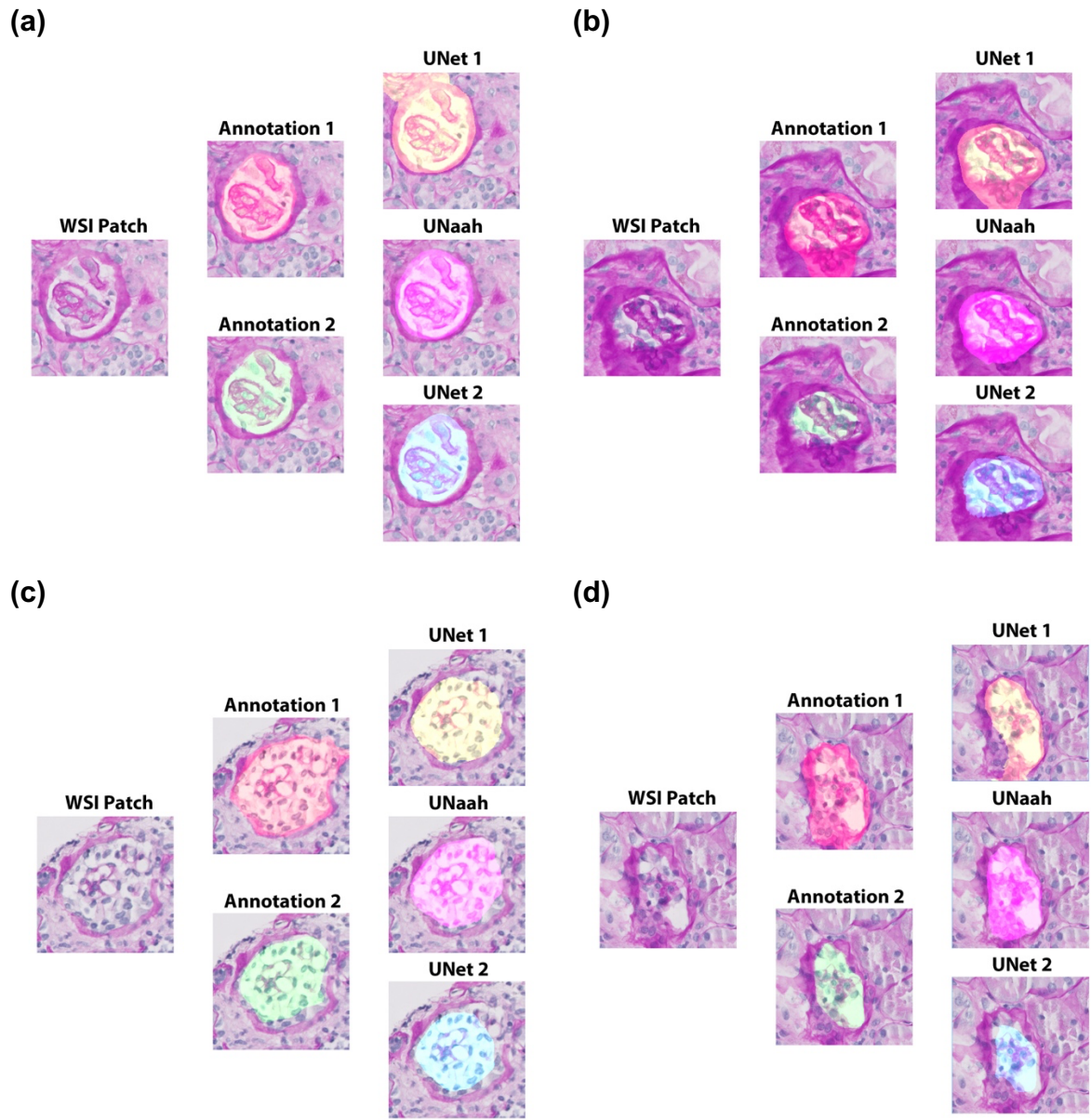

**Figure S2: Model performance on sample IVUS images.** In each sub-figure, the first column shows a representative IVUS image, and the second column shows annotations performed by the two radiologists highlighting the arterial cross section. The third column shows the performance of the UNet 1 on annotation 1, UNaah that learned from both the annotations and UNet 2 on annotation 2.

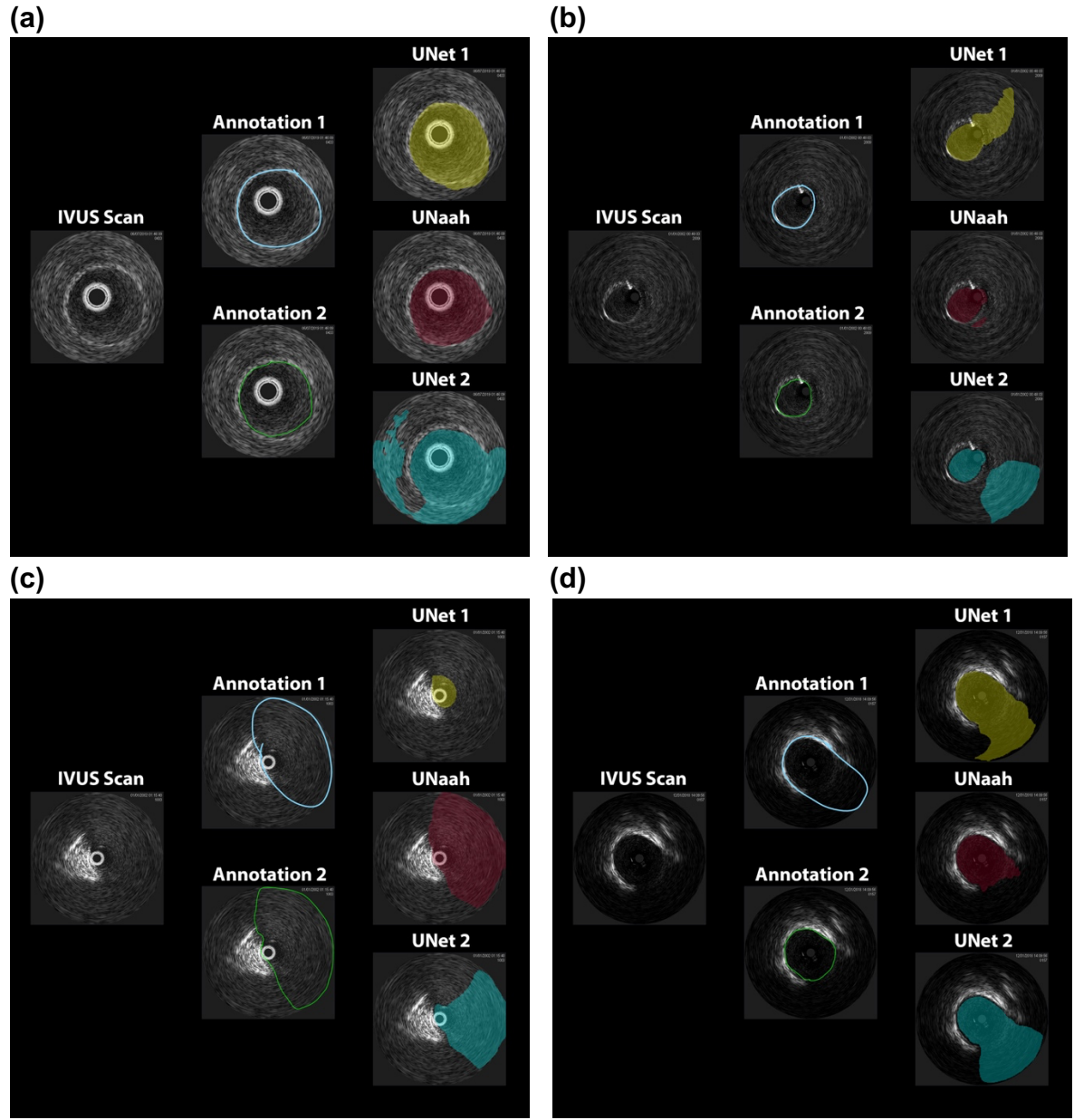